\documentclass{cernyrep} 
\usepackage{texnames}
\usepackage[T1]{fontenc}
\usepackage[bookmarks, colorlinks=true, linktoc=page, pdftex, linkcolor=black, citecolor=black, urlcolor=blue]{hyperref}

\usepackage{fancyhdr}
\fancyhfoffset{4 mm}
\fancypagestyle{ARTTITLE}{% 
\fancyhf{} % clear all header and footer fields
\lhead{\small{Proceedings of the 2018 CERN--Accelerator--School course on\\
\it{Numerical Methods for Analysis, Design and Modelling of Particle Accelerators}, Thessaloniki, (Greece)}} 
\lfoot{Available online at \url{https://cas.web.cern.ch/previous-schools}}
\rfoot{\thepage\hspace*{3mm}}
 
}
\renewcommand{\headheight}{22.38pt}

\begin{document}
\title{Monte Carlo Simulation Techniques}
 
\author {Ji Qiang}

\institute{Lawrence Berkeley National Laboratory, Berkeley, CA, USA}

\begin{abstract}
Monte Carlo simulations are widely used in many areas including
particle accelerators.
In this lecture, after a short introduction and reviewing of
some statistical backgrounds, we
will discuss methods such as direct inversion, rejection method, 
and Markov chain Monte Carlo to sample a probability distribution
function, and 
methods for variance reduction to evaluate numerical integrals
	using the Monte Carlo simulation.
We will also briefly introduce the quasi-Monte Carlo sampling at the end
of this lecture.
\end{abstract}


\maketitle % this produces the title block
 \thispagestyle{ARTTITLE}

\section{Introduction}
 
The Monte Carlo method is a (computational) method that relies on 
the use of random sampling and probability statistics to obtain 
numerical results for solving deterministic or probabilistic problems.
It is a method of solving various problems in computational mathematics by
constructing for each problem a random process with parameters equal to
the required quantities of that problem. The unknowns are determined 
approximately by carrying out observations on the random process and 
by computing its statistical characteristics which are approximately equal 
to the required parameters~\cite{halton}.

It is believed that the earliest documented use of random
sampling to solve a mathematical problem is by mathematician Comte de 
Buffon of France in 1777~\cite{kalos}. This problem is to find the probability of
an intersection between a randomly thrown needle of length L and
a group of parallel lines with separation width D. It turns out that the
analytical solution of this probability is proportional to $\pi$ and 
later suggested by Laplace to evaluate the $\pi$ using the random sampling.
Lord Kelvin used random sampling to aid in evaluating the time
integrals associated with the kinetic theory of gases and 
 Enrico Fermi was among the first to apply random sampling
methods to study neutron moderation in Rome.
During World War II,  Fermi, Stan Frankel, 
Nicholas Metropolis, John von Neumann, Stan Ulam and others developed computer-oriented 
Monte Carlo
methods at Los Alamos to study neutron transport through 
materials under the Manhattan project.
It is said that the name of "Monte Carlo" which is also a casino center for
gambling in Monaco, 
was coined by Metropolis
because of the similarity of the randomness employed in
the method and 
games of chance~\cite{wiki}.

The Monte Carlo simulation starts with a probability distribution function
that characterizes the parameters of the physical or mathematical system. Then one draws random sampling of the
distribution function to obtain a sample
of the parameters. Next, one runs simulation using those parameters. After that,
one collects the simulation outputs and repeats the above process for 
a number of samplings of the parameters. Finally, one
performs statistical analysis on the simulation outputs.
The Monte-Carlo simulation can be conveniently summarized in the following steps:
\begin{enumerate}
\item Define a domain of possible inputs and identify the statistical probability distribution of these inputs.
\item Generate possible inputs 
through random sampling from the probability distribution over the domain.
\item Perform simulation with these input parameters.
\item Aggregate and analyze statistically the output results.
\end{enumerate}
The error of the output results from the Monte Carlo simulation
is inversely proportional to the square root of the number of samples.

The Monte Carlo method can be used to solve
the problems that are stochastic (probabilistic) by nature such as
particle collision and transport, or the problems that 
are deterministic by nature such as the evaluation of integrals.
It has been used in areas as diverse as natural science such as physics and chemistry,
engineering such as control and optimization,
economics such as market prediction, and many others~\cite{kalos,wiki,mcphys,mcfe}.

\section{Statistical background}

In the Monte Carlo simulation, system parameters are treated
as random variables that follow 
some probability distributions.
The random variable is a real number associated with a random
event whose occurring chance is determined by an underlying
probability distribution. 
A discrete random variable such as face of dice or
type of reaction has a discrete probability distribution.
A continuous random variable such as spatial location or time of
occurrence has a continuous probability distribution.
If $x$ is a random variable 
with probability 
density function $p_i \delta(x-x_i)$ for the discrete variable and
$f(x)$ for the continuous variable, 
then $g(x)$ is also a random variable. 
The expectation of $g(x)$ is defined as:
\begin{eqnarray}
	E(g(x)) = <g(x)> & = & \sum_i p_i g(x_i); \ \ \mbox{for the discrete random variable} \\
	E(g(x)) = <g(x)> & = & \int_{-\infty}^{\infty} g(x) f(x) dx; \ \ \mbox{for the continuous random variable} 
\end{eqnarray}
where $p_i$ is the probability of the discrete random variable $x_i$, and
$f(x)$ is the probability density function (PDF) of the continuous variable
$x$.
The $n^{th}$ moment of random variable $x$ is defined as the expectation of the $n^{th}$ power of $x$.
The spread of the random variable is measured by
the variance of $x$. The square root of the variance is also
called standard deviation or standard error.
The variance of any function of the random variable is defined as:
\begin{equation}
	var(g(x)) = E(g^2(x)) - E^2(g(x))
\end{equation}
The variance has the following properties:
\begin{enumerate}
\item For a constant random variable $C$, $var\{C\} = 0$.
\item For a constant C and random variable $x$, $var\{Cx\} = C^2 var\{x\}$.
\item For independent random variables $x$ and $y$, $var\{x+y\} = var\{x\}+var\{y\}$
\end{enumerate}
When $x$ and $y$ are not necessarily independent, the covariance can be used to measure 
the degree of dependence of the two random variables $x$ and $y$:
\begin{equation}
	cov\{x,y\} = <xy> - <x><y>
\end{equation}
The covariance equals zero when $x$ and $y$ are independent and
\begin{equation}
	cov\{x,x\} = var\{ x \} 
\end{equation}
However,the zero covariance does not by itself guarantee independence of the random variables. 
For example, let $x$ be a uniform random variable between $-1$
and $1$, and let $y = x^4$, the covariance $cov\{x,y\} = 0$. 
%The covariance can have either a positive or negative value.
Another quantity to measure the dependence between two random variables 
is the correlation coefficient that is given by:
\begin{equation}
	\rho(x,y) = \frac{cov\{x,y\}}{var\{x\}var\{y\}}
\end{equation}
and 
\begin{equation}
	-1 \le \rho(x,y) \leq 1
\end{equation}

\section{Sampling of probability distribution function}
The Monte Carlo simulation starts with the sampling of a given probability distribution function.
In order to sample an arbitrary probability distribution, one needs first to
generate a uniformly distributed random number. The other complex probability distribution
can then be sampled based on this uniform random number through appropriate 
operations.

\subsection{Generation of uniformly distributed pseudo-random number}
On computer, instead of using a real random number, a pseudo-random
number is used to sample a uniform distribution between zero and one.
A simple and widely used algorithm to generate a pseudo-random 
number is called Linear Congruential Generator (LCG). The sequence
of numbers is given by the following recurrence relation:
\begin{equation}
	x_{k+1} = mod(a*x_k+c,M), \ \ k=1,2,\ldots
\end{equation}
where $x_{k+1}$ and $x_k$ are integers between 0 and $M$,
$mod$ is the module function, $M$ is the modulus, and $a$ and $c$ are the 
positive multiplier integer and the increment integer respectively.
This function has a largest period of $M$ if $M$, $a$, and $c$ are properly chosen,
and all possible integers between $0$ and $M-1$ can be attained starting
from an initial seed integer. Normally, $M$ is chosen as power of $2$ minus
one. This module function repeatedly brings the linear function $y=ax+c$ back to the range
between zero and $M$.
A uniformly distributed random number between $0$ and $1$ is given by:
\begin{equation}
	r = x_{k+1}/M 
\end{equation}
The typical choice for the $M$ is $2^{31}-1 = 2147483647$, $a=7^5=16807$,
and $c=0$.
%What's happening is we're drawing the line y = a * x + c "forever", but using the mod function (like wrap-around) to bring the line back into the square [0,m] x [0,m] 
%(m is power of 2 -1). By doing so, we've induced a map on the integers 0 through m which, if we've chosen a, c and m carefully, will do an almost perfect shuffle.
%A more ambitious LCG has the form:
%		SEED =(16807  SEED + 0) mod 2147483647
This is the random number generator that was used in function ran0 of the Numerical Recipe (NR)~\cite{nr}.
It shuffles the integers from 1 to 2,147,483,646, and then repeats itself.
However, there are serial correlations present in the above random number generator.
An improved version,
function ran1 of the NR, uses the function ran0 as its random value, 
but shuffles the output to remove low-order serial correlations. 
A random number derived from the $j^{th}$ value in the sequence, is 
not output on the $j^{th}$ call, but rather on a randomized later call, e.g. $j+32$ on average.
When a very long sequence of random number is needed,
one can
combine two different sequences with different periods so as to obtain a new sequence 
whose period is the least common multiple of the two periods. This is what implemented in
the function ran2 of the NR that has a period of ~$10^{18}$.

\subsection{Direct inversion}
The direct inversion method is also called transformation method.
The above section discussed how to generate 
a uniformly distributed random number between zero and one. Given the sampling 
of such a uniform probability density function, the sampling of the
other probability distribution function can be achieved through
appropriate transformation and inversion of that sampling. 
Given that $x$ is a random variable with probability density function $f(x)$ and $y = y(x)$, then
the probability density function of $y$ will be:
\begin{equation}
	g(y) = f(x)|\frac{dx}{dy}|
	\label{gy}
\end{equation}
which reflects the fact that all the values of $x$ in $dx$ map into values of $y$ in $dy$. 
In the above equation, one needs to use the function $x(y)=y^{-1}(x)$ to attain $g(y)$.
Consider the linear transformation $y = a + b\ x$,
the probability density function $g(y)$ will be:
\begin{equation}
	g(y) = f(\frac{y-a}{b})/|b|
\end{equation}
This suggests that in order to sample a Gaussian distribution with mean $\mu$ and 
standard deviation $\sigma$, one can sample a  
Gaussian distribution with mean zero and standard deviation one
and then transform the sampled variable $x$ using $y=\mu+\sigma x$.

The transformation Eq.~\ref{gy} can be rewritten in the integral form:
\begin{equation}
	\int_{-\infty}^y g(t) dt = \int_{-\infty}^x f(t) dt
\end{equation}
These integrals are called cumulative distribution functions (CDF)
of the random variable $y$ and $x$ respectively.
If $g(y)$ and $G(y)$ represent PDF and CDF of a random variable $y$, 
if a random number $x$ is distributed uniformly between zero and one
with PDF $f(x)=1$, the above equation can be rewritten as:
%if $y$ is such that
\begin{equation}
	G(y) = x
\end{equation}
Then for each uniformly distributed random variable
$x$, there is a corresponding $y$ that is 
distributed according to the probability density function $g(y)$.
For example, consider sampling a probability distribution function:
\begin{equation}
f(r) = r\exp(-\frac{1}{2}r^2), 0<r<\infty
\end{equation}
The cumulative distribution function of the above function is:
\begin{equation}
F(r) = \int_0^r t\exp(-\frac{1}{2}t^2) dt = 1-\exp(-\frac{1}{2}r^2) = \xi
\end{equation}
where $\xi$ is the uniformly distributed random variable with constant 
probability density function.
Solving this equation for $r$ yields:
\begin{equation}
	r = \sqrt{-2log(1-\xi)}
\end{equation}
Next, we would like to sample a Gaussian probability distribution function
with zero mean and standard deviation one:
\begin{equation}
	f(x) = \frac{1}{\sqrt{2 \pi}}\exp(-\frac{1}{2}x^2), \ \ -\infty<x<\infty
\end{equation}
We can construct a two-dimensional (2D) Gaussian probability density function:
\begin{equation}
	f(x,y) = \frac{1}{2 \pi}\exp(-\frac{1}{2}(x^2+y^2))
\end{equation}
Change the above coordinates $(x,y)$ into the cylindrical coordinates
$(r,\theta)$, the 2D probability distribution function 
becomes
\begin{equation}
	f(r,\theta) = \frac{1}{2 \pi}r \exp(-\frac{1}{2}r^2)
\end{equation}
This distribution can be sampled using the above example as:
\begin{eqnarray}
	\theta & = & 2 \pi \xi_1 \\
	r & = & \sqrt{-2\log(1-\xi_2)}
\end{eqnarray}
then in the Cartesian coordinate:
\begin{eqnarray}
	x & = & r \cos(\theta) \\
	y & = & r \sin(\theta)
\end{eqnarray}
The above equations can be rewritten as:
\begin{eqnarray}
	x & = & \sqrt{-2\log \xi_2} \cos(2\pi \xi_1) \\
	y & = & \sqrt{-2\log \xi_2} \sin(2\pi \xi_1) 
\end{eqnarray}
Here, the uniform random number $\xi_2$ is used to replace the 
original uniform random number $1-\xi_2$.
The above sampling of a Gaussian distribution function is known as
the Box-Muller method~\cite{box}.

For a complex probability distribution function whose CDF is not
analytically available, one can numerically calculate a discrete CDF as:
\begin{equation}
	F(x_n) = \int_0^{x_n} f(t) dt = \frac{n}{N}, \ \ n=0,1,2,\ldots,N
\end{equation}
For a uniformly sampled random number $\xi$ from $u(0,1)$,
one can find $n$ such that:
\begin{equation}
	\frac{n}{N} < \xi <\frac{n+1}{N}
\end{equation}
The sampled value for $x$ can be calculated by the following linear interpolation:
\begin{eqnarray}
	x & = & x_n + (x_{n+1}-x_n) r \\
        r & = & N \xi - n, \ \ 0 < r< 1	
\end{eqnarray}

For a discrete probability distribution function,
$f(x) = p_i\delta(x-x_i)$ and $\sum_i p_i = 1$, $i=1,2,\ldots,N$,
one can generate a uniform random number $\xi$, and obtain
the sampled random variable $x=x_k$ so that
\begin{equation}
\sum_{i=1}^{k-1} p_i \leq \xi < \sum_{i=1}^k p_i
\end{equation}

For a multi-dimensional probability distribution function,
if the random variable in each dimension is independent of each other, 
the sampling of multi-dimensional PDF can be done in each dimension separately.
If the marginal and the conditional functions can be determined, 
sampling the multivariate distribution will then involve sampling the sequence of univariate distributions.

\subsection{Rejection method}
\begin{figure}
\centering\includegraphics[width=.5\linewidth]{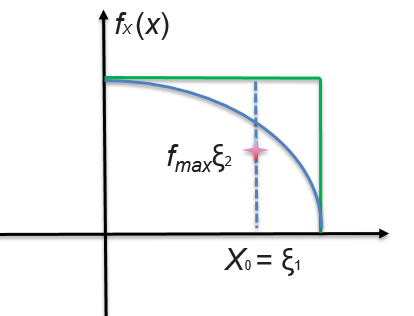}
\caption{A geometric view of the rejection method.}
\label{mcrej}
\end{figure}

In many applications, the multi-dimensional probability distribution function 
can be very complicated and the 
explicit analytical expression of the cumulative distribution function is not
attainable. The above direct inversion of the CDF becomes impossible. 
In this case, the rejection method can be used as a general method to sample
the probability distribution function.
The rejection method is a composition method that needs two samplings to sample
a given distribution. Here, the first sampling will generate a random point within the
variable domain of the probability distribution function. The second sampling 
will generate a uniform random number between zero and one. The probability
of accepting the first sampling point depends on 
the normalized function value at the first sampling point. If the 
uniform random number is less than or equal to the normalized function value,
the first sampling point is accepted as the sampling point of the probability
distribution function, otherwise, it is rejected.
For a one-dimensional PDF, the rejection method can be written as follows:
\begin{itemize}
\item generate a uniform random number $x_0=\xi_1$ between $x_{min}$ and $x_{max}$.
\item generate another uniform random number $\xi_2$ between 0 and 1. 
\item if $\xi_2 \leq \frac{f(x_0)}{f_{max}}$: accept $x_0$. 
\item otherwise; reject $x_0$.
\end{itemize}
Here, $f_{max}$ is the maximum value of the PDF within the domain
between $x_{min}$ and $x_{max}$.
A geometric view of the rejection method is shown in Fig.~\ref{mcrej}.
Here, the rejection method can be viewed as to choose
uniformly the points enclosed by the curve $f(x)$ inside the smallest
rectangle that contains the curve. The ordinate of such a point is 
$x_0 = \xi_1$; the abscissa is $f_{max}\xi_2$. Points lying above the curve 
are rejected; points below are accepted. Their ordinates $x = x_0$ 
have the distribution $f(x)$.
For example, consider the following probability
distribution function:
\begin{equation}
	f(x) = \frac{1}{1+2x^2}, 0<x<1
\end{equation}
This can be sampled using the following steps:
\begin{enumerate}
\item $x_0 = \xi_1$
\item if $\xi_2 > \frac{1}{1+2x_0^2}$, repeat from 1; else $x=x_0$.
\end{enumerate}
where $\xi_1$ and $\xi_2$ are two uniformly sampled 
random numbers between zero and one.
Another example of using the rejection method is to sample
a uniform density distribution within a unit circle. 
This can be done as follows:
\begin{enumerate}
\item $x_0 = \xi_1$, and $y_0 = \xi_2$
\item if $x_0^2+y_0^2 > 1$, repeat from 1; else $x=x_0$ and $y=y_0$.
\end{enumerate}

The above rejection method requires the information of the maximum value
of the sampled probability distribution function
within the domain in order to calculate the
normalized function value $\frac{f(x_0)}{f_{max}}$. 
The efficiency of the rejection method depends on the ratio of $f(x_0)/f_{max}$. 
In many applications, 
this ratio can be small, e.g. the tail
of a Gaussian distribution. This suggests that many trial solutions will
be rejected before attaining a sampled point.
For some complex probability distribution function, the maximum of the 
function is not easily accessible. However, if one can find an easily sampled
function $g(x)$
so that $M g(x) \ge f(x)$ with constant $M>1$ for all $x$, 
a general rejection method can be written as~\cite{wikirej}:
\begin{itemize}
\item generate a random number $x_0$ between $x_{min}$ and $x_{max}$ from the sampling of $g(x)$.
\item generate a uniform random number $\xi_1$ between $0$ and $1$. 
\item if $\xi_1 \leq f(x_0)/(Mg(x_0))$: accept $x_0$. 
\item otherwise; reject $x_0$.
\end{itemize}
If one choose the $g(x) = f_{max}/M$, a uniform distribution, the
above general rejection method is reduced to the preceding rejection method.
It is clear that the efficiency of the above rejection method depends on the
ratio of $f(x)/Mg(x)$.

\subsection{Markov chain Monte Carlo}
The efficiency of the rejection method can be improved by another general
sampling method, Metropolis method or in general also called 
Markov chain Monte Carlo (MCMC) method. 
The Markov chain Monte Carlo method is a general method to sample 
any probability distribution function regardless of its analytic complexity
in any number of dimensions.
It does not need to know either the maximum or the upper bound of the sampled 
probability distribution function. Moreover, it does not reject all samplings
with $f(x_0)/f_{max} < \xi_2$ in the rejection method but with a probability of acceptance depending on
the local function value. This makes it more efficient than
the rejection method. Some
disadvantages of the MCMC are that the sampling is correct only asymptotically 
and that successive samplings are correlated.
To avoid these disadvantages, some initial samplings are thrown away
(called burn-in phase) and the used samplings are separated by a number of 
steps.

A sequence of random variables $x_i, i = 1, 2, \ldots$ forms a 
Markov chain if:
\begin{equation}
	P(x_{i+1} = x|x_1,\cdots,x_i) = P(x_{i+1} = x|x_i)
\end{equation}
That is, the probability distribution of $x_{i+1}$ depends only on the
previous step $x_i$, and is independent of other
steps ($x_{i-1},\ldots,x_1$) before.
A Markov chain is said to be ergodic if it satisfies the
following conditions~\cite{wikimarkov,lam,breheny}:
\begin{itemize}
\item Irreducible: Any state can be reached from any other state with nonzero probability.
\item Positive recurrent: For any state A, the expected number of steps required for the chain to return to A is finite.
\item Aperiodic: For any state A, the number of steps required to return to A must not always be a multiple of 
some integer value.
\end{itemize}
In other words, it means that all possible states of
the system will be reached within some finite number of steps.
If there exists a distribution function $f(x)$ such
that $f(x_{i+1})P(x_{i+1}|x_i) = f(x_i)P(x_i|x_{i+1})$ for all $i$,
the Markov chain is reversible and the $f(x)$ is then the
equilibrium distribution of the Markov chain.
Provided that a Markov chain is ergodic it will converge to an
equilibrium stationary distribution. 
%also called the
%stationary distribution.
This stationary distribution is determined entirely by the
transition probabilities of the chain. The initial value of the
chain is irrelevant in the long run. This suggests that
the sampling based on the Markov chain would sample the 
probability distribution $f(x)$ asymptotically.
The Metropolis Markov chain Monte-Carlo algorithm to sample
an arbitrary distribution can be summarized as:
\begin{enumerate}
\item choose a proposal transition probability distribution $p(x)$ and
an initial random sampled value $x_1$.
\item calculate a new trial value ${\bar x} = x_i + \tau$
using an update step $\tau$ sampled from $p(x)$.
\item if $f(\bar x) > f(x_i)$ accept $x_{i+1} = \bar{x}$, otherwise
accept $x_{i+1} = \bar{x}$ with a probability $f({\bar x})/f(x_i)$.
\item continue step 2 until one has enough number of sampled values.
\item discard some early values during the burn-in phase.
\end{enumerate}
Typically, the proposal distribution function can be
assumed as a Gaussian function $p(x) = \mathcal{N}(0,\sigma)$ or
a uniform distribution $p(x)=\mathcal{U}(-v,v)$~\cite{paltani}.

The above symmetric proposal transition distribution might not be optimal.
In order to speed up convergence, a correction factor, the
Hastings ratio, is applied to correct for the bias.
The probability to accept the new trial sampling value is modified
from the original $min(1,f(\bar{x}/f(x_i))$ to include the
Hastings ratio $min(1,f(\bar{x})p({\bar x};x_i)/(f(x_i) p(x_i;\bar{x}))$.
If $p({\bar x};x_i)=p(x_i;{\bar x})$, 
this is the Metropolis algorithm.

\begin{figure}
\centering\includegraphics[width=.5\linewidth]{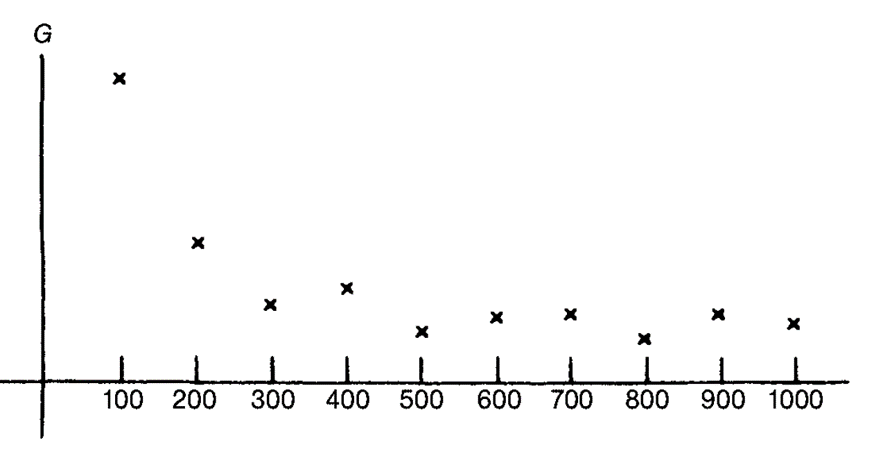}
	\caption{Behaviour of <G> with length of the $M(RT)^2$ random walk~\cite{kalos}.}
\label{mcmc}
\end{figure}

In practical application, the width of the proposal distribution 
(e.g. for a Gaussian update or for a uniform update) should be tuned 
during the burn-in phase to set the rejection fraction in the right range.
The conventional acceptance probability is typically
between $30\%$ and $70\%$.
One can use the autocorrelation function to check if the initial value
has become irrelevant or not.
In order to break the dependence between successive draws in the Markov chain, 
one might keep only every $n^{th}$ draw of the chain.
To check whether a Markov chain reaches an equilibrium distribution or not,
one can use multiple chains. 
When the variance between multiple chains is much less than the variance within the chains, the chain reaches the equilibrium.
One can also monitor the behaviour of an expectation value that evolves with the length of the
Markov chain random walk. 
Figure~\ref{mcmc} shows the evolution of the expectation value calculated from
$100$ samplings as a function of the random walk steps.
It is seen that after $400$~random walks, the expectation value starts to fluctuate.
This suggests that the chain might have reached an equilibrium state.

\section{Numerical integration using the Monte Carlo method}

One of the most important applications of the Monte Carlo method is to
calculate the integral.
Given the following integral:
\begin{equation}
	G = \int g(x)f(x) dx, \ \ f(x)\ge 0 \ and \ \int f(x) = 1
\label{integ}
\end{equation}
one can sample the probability distribution function $f(x)$ and
form the arithmetic mean as:
\begin{equation}
	G_N = \frac{1}{N} \sum_i g(x_i)
\end{equation}
where $N$ is the number of sampling points.
The original integration can be written as:
\begin{equation}
	G = G_N + error 
\end{equation}
with
\begin{equation}
	|error| \simeq \frac{\sigma}{\sqrt{N}} 
\label{error}
\end{equation}
where 
\begin{equation}
	\sigma^2 = \int g^2(x)f(x)dx - G^2
\end{equation}
denotes the variance of the function $g(x)$.
The error will decrease as $1/\sqrt{N}$ {\bf independent of dimensionality} 
of the integral. 
This is the key advantage of the Monte Carlo method
over the direct numerical quadrature whose computational cost scales {\bf exponentially with the 
number of dimensions}. In order to reduce the error in the calculation
of the integral using the Monte Carlo method, 
for a given number of samplings, one needs to reduce
the variance of the integrand or to improve on the scaling
with respect to the number of samplings.
In the following, we will introduce several variance reduction methods
and a quasi-Monte Carlo method to reduce the numerical error in the
evaluation of the integral using the Monte Carlo method.

\subsection{Importance sampling for variance reduction}
Given the initial integral Eq.~\ref{integ}, we can rewrite the integral as:
\begin{equation}
	G = \int \frac{g(x)f(x)}{\bar f(x)}{\bar f}(x)dx
\end{equation}
where $\bar f(x)$ is a new probability density function.
Using the sampling from this new probability distribution function,
the numerical integral can be written as:
\begin{equation}
	G_N = \frac{1}{N} \sum_i \frac{g(x_i)f(x_i)}{\bar f(x_i)}
\end{equation}
The variance of the new integrand becomes:
\begin{equation}
	\bar {\sigma}^2 = \int \frac{g^2(x)f^2(x)}{\bar f(x)}dx - G^2
\end{equation}
Ideally, the optimal $\bar f(x)$ should be chosen as $g(x) f(x)/G$~\cite{kalos}. 
%the value of $G$ is not known. 
In practice, a similar function
to the integrand $g(x)f(x)$ can be used as $\bar f(x)$ to reduce the
variance.
For example, consider the following integral:
\begin{equation}
	G = \int_0^1 \cos(x)dx
\end{equation}
A straightforward Monte Carlo algorithm would be to sample a
uniform probability density function $f(x)$ between [0, 1], 
and then to calculate the mean quantity $<\cos(x)>$. The variance of
this direct Monte Carlo calculation is $0.01925$.
If we approximate the original function as:
\begin{equation}
	\cos(x) \approx 1 - \frac{x^2}{2} 
\end{equation}
and choose ${\bar f}(x) = \frac{6}{5}(1-x^2/2)$ as the importance sampling function,
the new function will be:
\begin{equation}
	{\bar g}(x) = \frac{\cos(x)}{{\bar f}(x)}=\frac{10}{6} \frac{\cos(x)}{2-x^2}
\end{equation}
The variance of the new function is $0.0001949$. This is about two
orders of magnitude reduction of the variance 
in comparison to the direct Monte-Carlo method.

\subsection{Correlation methods for variance reduction}

Besides the importance sampling method to reduce the variance,
another way to reduce the variance is to make use some function
whose integral can be calculated analytically. 
Using the analytically integrable function $h(x)$,
the original integral Eq.~\ref{integ} can be rewritten as:
\begin{equation}
	G = \int (g(x)-h(x))f(x) dx + \int h(x)f(x) dx
\end{equation}
Here, we assume that the second integral can be obtained
analytically. Using the Monte Carlo method to sample
the probability distribution $f(x)$, the above integral can be 
approximated as:
\begin{equation}
	G \simeq \frac{1}{N} \sum_i(g(x_i)-h(x_i)) + \int h(x)f(x) dx
\end{equation}
If the variance of $g(x)-h(x)$,
is much less than the variance
of the original function $g(x)$, especially, if
$|g(x)-h(x)|$ is approximately constant for different values
of $h(x)$, then the above
correlated sampling would be an efficient variance reduction method.
For example, consider the following integral:
\begin{equation}
	G = \int_0^1 \sin(x) dx
\end{equation}
The variance of $G$ using the direct Monte Carlo method following a
uniform probability distribution function is $0.0614$.
If we choose $h(x) = x$, the variance of $\sin(x) -x $ following
the same uniform probability distribution is $0.00205$, which
is more than an order of magnitude less than the variance of the original function.

\subsection{Method of antithetic variates}
This method exploits the fact that the decrease in variance occurs when random variables 
are negatively correlated. 
Consider the following integral:
\begin{equation}
	G = \int_0^1 g(x) dx 
\end{equation}
This integral can be rewritten as:
\begin{equation}
	G = \int_0^1 \frac{1}{2}[g(x) + g(1-x)] dx 
\end{equation}
and
\begin{equation}
G_N = \frac{1}{N} \sum_i[g(x_i)+g(1-x_i)] 
\end{equation}
If $g(x)$ is a linear function of $x$, the variance of the
above integration will be zero. 
For nearly
linear functions, this method will substantially reduce the variance.
For example, consider the following integral:
\begin{equation}
	G = \int_0^1 e^{-x} dx 
\end{equation}
The variance using the direct Monte Carlo method (assuming a uniform
density function $f(x)$) is $0.019$. Using the above method,
the variance is reduced to $0.00053$, another order of magnitude 
reduction of the variance.

\subsection{Quasi-Monte Carlo non-random sampling}
\begin{figure}
\centering\includegraphics[width=.45\linewidth]{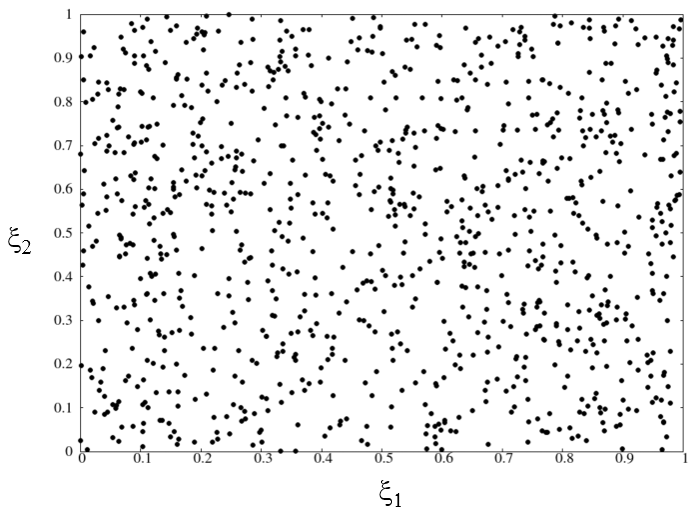}
\centering\includegraphics[width=.45\linewidth]{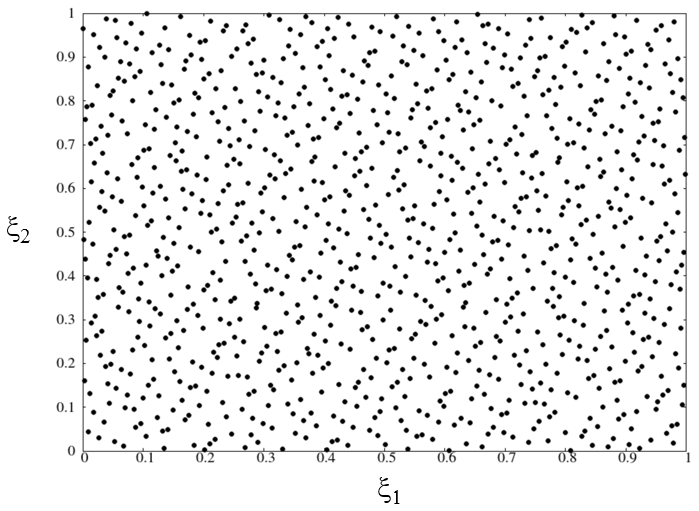}
\caption{Two dimensional uniform sampling from the Fortran pseudo-random number generator (left)
	and the Halton sequence (right).}
\label{mcsamp}
\end{figure}

As seen from the Eq.~\ref{error}, in order to reduce the error in numerical
integration using the Monte Carlo method, besides reducing the
variance of integral using the preceding methods, another way 
is 
%to increase the number of samplings or 
to improve on the scaling with respect to the number of samplings.
A quasi-Monte Carlo sampling is a method that uses a non-random sequence
to sample the uniform distribution between zero and one.
Sampling an arbitrary probability distribution can then be attained through the 
transformation of the sampling of the uniform distribution.
A non-random sequence that has low discrepancy (a measure of deviation from uniformity) 
can be used to simulate the uniform distribution. A popular non-random
Halton/Hammersley sequence in multiple dimensions is defined as 
follows~\cite{lavalle}:
\begin{eqnarray}
X & = & {(j-1/2)/N,\Phi_2(j),\Phi_3(j),\ldots,\Phi_r(j)}, \ \ j=1,\ldots,N \\
j & = & a_0 + a_1 r^1 + \ldots \\
\Phi_r(j) & = & a_0 r^{-1} + a_1 r^{-2}+\ldots 
\end{eqnarray}
where $\Phi_r(j)$ is the radical inversion function in the base of a prime
number $r$. For example, using base number $3$, and $j=1,2,3,4$
one obtains the sequence: $\Phi_3(1) = 1/3$, 
 $\Phi_3(2) = 2/3$,  $\Phi_3(3) = 1/9$,  $\Phi_3(4) = 4/9$. 
Figure~\ref{mcsamp} shows $1000$ samplings of a two-dimensional uniform distribution
from using the random sampling and from the non-random Halton sequence.
It is seen that the non-random sampling populates the two-dimensional square
more uniformly than the random sampling.
Fluctuation of this type of sequence scales as $1/N$ whereas a random 
Monte Carlo sampling scales as $1/\sqrt{N}$.
The error in some cases of numerical integration using the non-random sampling
can reach $1/N$~\cite{nr}.

%\begin{figure}
%\centering\includegraphics[width=.5\linewidth]{intqsamp.png}
%\caption{Description of my figure}
%\label{mcrej}
%\end{figure}

\section*{Acknowledgements}
This work was supported by the U.S. Department of Energy under Contract No. DE-AC02-05CH11231.

\end{document}